\documentclass{ws-ijmpa}
\def \bea{\begin{eqnarray}}
\def \beq{\begin{equation}}
\def \eea{\end{eqnarray}}
\def \eeq{\end{equation}}
\def \b{{\cal B}}
\def \bo{B^0}

\def \ob{\overline{B}^0}

\def \cpp{C_{+-}}

\def \lpp{\lambda_{\pi \pi}}
\def \ob{\overline{B}^0}

\def \s{\sqrt{2}}

\def \3half{\frac{3}{2}}
\def \spp{S_{+-}}
\renewcommand{\thetable}{\Roman{table}}
\begin{document}

\markboth{M. Gronau}
{CP violation in beauty decays}

\catchline{}{}{}{}{}

\title{CP VIOLATION IN BEAUTY DECAYS\footnote{Based partially on review talks given at 
recent conferences.}}

\author{\footnotesize MICHAEL GRONAU}

\address{Physics Department, Technion -- Israel Institute of Technology\\
32000 Haifa, Israel\\
gronau@physics.technion.ac.il}

\maketitle


\begin{abstract}
Precision tests of the Kobayashi-Maskawa model of 
CP violation are discussed, pointing out possible 
signatures for other sources of CP violation and for new flavor-changing operators. 
The current status of the most accurate tests is summarized.

\end{abstract}

\ccode{PACS Nos.: 13.25.Hw, 11.30.Er, 12.15.Ji, 14.40.Nd}

\section{Introduction}	

It took thirty-seven years from the discovery
of a tiny  CP violating effect of order $10^{-3}$ in $K_L\to\pi^+\pi^-$~\cite{Christenson:1964fg}  
to a first observation of a breakdown of CP symmetry outside the strange 
meson system.  A large CP asymmetry of order one between rates of
initial $B^0$ and $\bar B^0$ decays to $J/\psi K_S$ was measured  in summer 2001 by the 
Babar and Belle Collaborations.\cite{Aubert:2001nu} A sizable however smaller 
asymmetry had been anticipated twenty years earlier\,\cite{Carter:1980hr} in the  framework 
of the Kobayashi-Maskawa (KM) model of CP violation,\cite{Kobayashi:1973fv} in the absence
of crucial information on $b$ quark couplings.  The asymmetry was observed 
in a time-dependent measurement as suggested,\cite{Dunietz:1986vi}
thanks to the long $B^0$ lifetime and the large $B^0$-$\bar B^0$ 
mixing.\cite{Albrecht:1987dr} The measured 
asymmetry, fixing (in the 
standard phase convention\cite{Wolfenstein:1983yz}) the sine of the phase 
$2\beta~(\equiv 2\phi_1) \equiv 2{\rm arg}(V_{tb}V^*_{td})$ of the top-quark dominated 
$B^0$-$\bar B^0$ mixing amplitude,  was found to be in good agreement 
with other determinations of Cabibbo-Kobayasi-Maskawa (CKM) 
parameters,\cite{Charles:2006yw,Bona:2006ah} including a recent precise measurement 
of $B_s$-$\bar B_s$ mixing.\cite{Abazov:2006dm} This showed that
the CKM phase  $\gamma~(\equiv \phi_3) \equiv {\rm arg}(V^*_{ub})$, which seems to be 
unable to 
account for the observed cosmological baryon asymmetry,\cite{Dolgov:2005wf}
is the dominant source of CP violation in flavor-changing processes. 
With this confirmation the next pressing question became whether small contributions 
beyond the CKM framework occur  in CP violating flavor-changing 
processes, and whether such effects can be observed in beauty decays. 
 
One way of answering this question is by over-constraining the CKM unitarity triangle
through precise CP conserving measurements related to the lengths of the sides of the triangle. 
An alternative and more direct way, focusing on the origin of CP violation in the CKM framework,  
is to measure $\beta$ and $\gamma$ in a variety of $B$ decay modes. 
Different values obtained  
from asymmetries in several processes, or values different from those imposed  
by other constraints, could provide clues for new sources of CP violation and for new 
flavor-changing interactions. Such phases and interactions occur in the low energy 
effective Hamiltonian of extensions of the Standard Model (SM) including models based on 
supersymmetry.\cite{Gabrielli:1995bd} 

In this presentation we will focus on the latter approach based primarily on CP asymmetries, 
using also complementary information on hadronic $B$ decay rates which are expected to be related 
to each other in the CKM framework. In the next section we outline several of the most relevant 
processes and the theoretical tools applied for their studies, quoting numerous papers 
where these ideas have been originally proposed and where more details can be 
found.\cite{References} Sections 3, 4 and 5 describe a number of methods in some 
detail, summarizing at the end of each section the current experimental situation. Section 6 
discusses several tests for NP effects, while Section 7 concludes. 

\section{Processes, methods and New Physics effects} 

Whereas testing the KM origin of CP violation in most hadronic $B$ decays requires
separating strong and weak interaction effects, in a few ``golden modes"  
CP asymmetries are unaffected by strong interactions. For instance, the decay 
$B^0\to J/\psi K_S$ is dominated by a single tree-level quark transition $\bar b\to \bar c c \bar s$,
up to a correction smaller than a fraction of a 
percent.\cite{Gronau:1989ia,Boos:2004xp,Ciuchini:2005mg,Li:2006vq} Thus, the asymmetries 
measured in  this process and in other decays dominated by $\bar b\to \bar c c \bar s$ have 
already provided a rather precise measurement of 
$\sin 2\beta$,\cite{Aubert:2006aq,Chen:2006nk,HFAG}
\beq
\sin 2\beta = 0.678 \pm 0.025~.
\eeq 
This value permits two solutions for $\beta$ at $21.3^\circ$ and at $68.7^\circ$.
Time-dependent angular studies of $B^0\to J/\psi K^{*0}$,\cite{Aubert:2004cp} and time-dependent Dalitz analyses of $B^0 \to Dh^0~(D\to K_S\pi^+\pi^-, h^0=\pi^0, \eta, \omega)$\cite{Krokovny:2006sv}
measuring $\cos 2\beta > 0$ have excluded the second solution at a high confidence level, implying
\beq
\beta = (21.3 \pm 1.0)^\circ~.
\eeq
Since $B^0\to J/\psi K_S$ proceeds through a CKM-favored quark transition, contributions to the 
decay amplitude from physics at a higher scale  are expected to be very small,  
potentially identifiable by a tiny direct asymmetry in this process or in 
$B^+\to J/\psi K^+$.\cite{Fleischer:2001cw}  

Another process where the determination of a weak phase is not affected
by strong interactions is $B^+\to DK^+$, proceeding through tree-level amplitudes
$\bar b\to \bar c u \bar s$ and $\bar b\to \bar u c \bar s$. The interference 
of these two amplitudes,  from $\bar D^0$ and $D^0$ which can always decay 
to a common hadronic final state, leads to decay rates and a CP asymmetry which measure 
very cleanly the relative phase $\gamma$ between these 
amplitudes.\cite{Gronau:1990ra,Gronau:1991dp} 
The trick here lies in recognizing the measurements which yield this fundamental CP-violating 
quantity. Physics beyond the SM is expected to have a negligible effect on  
this determination of $\gamma$ which relies on the interference of two tree amplitudes.
   
$B$ decays into pairs of charmless mesons, such as $B\to \pi\pi$ (or $B\to \rho\rho$) and 
$B\to K\pi$ (or $B\to K^*\rho$),  involve contributions of both tree and penguin amplitudes 
which carry different weak and strong phases.\cite{Gronau:1989ia,London:1989ph,Grinstein:1989df} 
Contrary to the case of $B\to DK$, the determination of $\beta$ and $\gamma$ using CP 
asymmetries in charmless $B$ decays involves two correlated aspects which must be considered: 
its dependence on strong interaction dynamics and its sensitivity to potential New 
Physics (NP) effects.  
This sensitivity follows from the CKM and  loop suppression of penguin amplitudes, implying 
that new heavy particles at the TeV mass range, replacing the $W$ boson and the top-quark in 
the penguin loop, may have sizable effects.\cite{Gronau:1996rv}. In order to claim evidence for
physics beyond the SM from a determination of $\beta$ and $\gamma$ in these processes
one must handle first the question of dynamics. There are two approaches for treating the 
dynamics of charmless hadronic $B$ decays:

\medskip
(1) Study systematically strong interaction effects in the framework of QCD.

(2)  Identify by symmetry observables which do not depend on QCD dynamics.

\medskip
The first approach faces the difficulty of having to treat precisely long distance 
effects of QCD including final state interactions.  Remarkable theoretical progress has been made
recently in proving a leading-order (in $1/m_b$) factorization formula for these amplitudes in 
a heavy quark effective theory approach to perturbative 
QCD.\cite{Beneke:1999br,Keum:2000ph,Bauer:2004tj} However, there remain differences between 
ways of treating in different approaches  power counting, the scale of Wilson coefficients, 
end-point quark distribution functions of  light mesons, and nonperturbative contributions from charm loops.\cite{Ciuchini:1997hb} 
Also,  the nonperturbative input parameters in these calculations involve non-negligible uncertainties. 
These parameters include heavy-to-light form factors at small momentum transfer, light-cone 
distribution amplitudes, and the average inverse momentum fraction of the spectator quark in 
the $B$ meson. The resulting inaccuracies in calculating magnitudes and strong phases of 
amplitudes prohibit a precise determination of $\gamma$ from measured decay rates and CP
asymmetries. Also, the calculated rates and asymmetries cannot provide a clear case for physics 
beyond the SM in cases where the results of a calculation deviate slightly from the measurements.

In the second approach one applies isospin symmetry to obtain relations among 
several decay amplitudes.  For instance, using the distinct behavior under isospin of tree 
and penguin operators contributing in $B\to\pi\pi$, a judicious choice of observables 
permits a determination of $\gamma$ or $\alpha~(\equiv \phi_2) = \pi-\beta -
\gamma$.~\cite{Gronau:1990ka}
The same analysis applies in $B$ decays to pairs of longitudinally polarized $\rho$ mesons.
In case that an observable related to the subdominant penguin amplitude is not measured 
with sufficient  precision, it may be replaced in the analysis by a CKM-enhanced SU(3)-related observable, in which a large theoretical uncertainty is translated to a small error in $\gamma$.
The precision of this method is increased by including contributions of higher order
electroweak penguin amplitudes, which are related by isospin to tree 
amplitudes.\cite{Buras:1998rb,Gronau:1998fn}  With sufficient statistics one should 
also take into account isospin-breaking corrections
of order $(m_d - m_u)/\Lambda_{\rm QCD} \sim 0.02$,\cite{Gardner:1998gz,Gronau:2005pq}
and an effect caused by the $\rho$ meson width.\cite{Falk:2003uq}
A similar analysis proposed for extracting $\gamma$ in 
$B\to K\pi$~\cite{Nir:1991cu,Deshpande:1994pw} requires using flavor SU(3)  instead of 
isospin for relating electroweak penguin contributions and tree 
amplitudes.\cite{Gronau:1998fn,Neubert:1998pt} While flavor SU(3) is 
usually assumed to be broken by corrections of order $(m_s-m_d)/\Lambda_{\rm QCD}\sim 0.3$,
in this particular case a rather precise recipe for SU(3) breaking is provided by QCD factorization, 
reducing the theoretical uncertainty in $\gamma$ to only a few degrees.\cite{Neubert:1998re} 

Charmless $B$ decays, which are sensitive to physics beyond the SM~\cite{Gronau:1996rv}, 
provide a rich laboratory for studying various signatures of NP.  A large variety of theories 
have been studied in this context, including supersymmetric 
models, models involving tree-level flavor-changing $Z$ or $Z'$ couplings, models with 
anomalous three-gauge-boson couplings and other models involving an enhanced 
chromomagnetic dipole operator.\cite{Grossman:1996ke,Ciuchini:1997zp}
The following effects have been studied and will be discussed in Section 6
in a model-independent manner:
\vskip 1mm
(1) Within the SM, the three values of $\gamma$ extracted from $B\to\pi\pi$, $B\to K\pi$ 
and $B^+\to DK^+$ are equal. As we will explain, these three values are expected to be different
in extensions of the SM involving  new low energy  four-fermion operators behaving 
as $\Delta  I=3/2$ in $B\to\pi\pi$ and  as $\Delta I = 1$ in $B\to K\pi$. 
\vskip 1mm
(2) Other signatures of anomalously large  $\Delta I = 1$ operators contributing to 
$B\to K\pi$ are violations of isospin sum rules, holding in the SM for both decay rates 
and CP asymmetries in these decays.\cite{Gronau:1998ep,Atwood:1997iw,Gronau:2005gz}
\vskip 1mm
(3)  Time-dependent asymmetries in $B^0 \to \pi^0K_S$, $B^0\to \phi K_S$ 
and $B^0\to \eta'K_S$ and in other $b\to s$ penguin-dominated decays may differ
substantially from the asymmetry $\sin 2\beta\sin\Delta mt$, predicted approximately in 
the SM.\cite{London:1989ph,Grossman:1996ke,London:1997zk} Significant deviations are 
expected in models involving anomalous $|\Delta S|=1$ operators behaving as 
$\Delta I=0$ or $\Delta I=1$. 
\vskip 1mm
(4) An interesting question, which may provide a clue to the underlying New Physics 
once deviations from SM predictions are observed, is how 
to diagnose the value of $\Delta I$ in NP operators contributing to $|\Delta S|=1$ 
charmless $B$ decays. We will discuss an answer to this question which 
has been proposed  recently.\cite{Gronau:2007ut}

\section{Determining $\gamma$ in $B\to DK$}

In this section we will discuss in some length a rather rich and very precise method for 
determining $\gamma$ in processes of the form $B\to D^{(*)}K^{(*)}$, which uses both 
charged and neutral $B$ mesons and a large variety of final states. It is based on a 
broad idea
that any coherent admixture of a state involving a $\bar D^0$ from $\bar b\to \bar cu \bar s$ 
and a state with $D^0$ from $\bar b \to \bar uc\bar s$ can decay to a common final 
state.\cite{Gronau:1990ra,Gronau:1991dp} The interference between the two 
channels, $B\to D^{(*)0}K^{(*)},~D^0\to f_D$ and $B\to \bar D^{(*)0}K^{(*)},~\bar D^0\to f_D$, 
involves the weak phase difference $\gamma$, which 
may be determined with a high theoretical precision using a suitable choice of measurements. 
Effects of $D^0$-$\bar D^0$ mixing are negligible.\cite{Grossman:2005rp} 
While some of these processes are statistically limited, combining them together is expected to reduce 
the experimental error in $\gamma$. In addition to (quasi) two-body $B$ decays, the $D$ or $D^*$ 
in the final state may be accompanied by any multi-body 
final state with quantum numbers of  a kaon.\cite{Gronau:1991dp}

Each process in this large class of neutral and charged $B$ decays is characterized by two pairs 
of parameters, describing complex ratios of amplitudes for $D^0$ and $\bar D^0$ for the two steps 
of the decay chain (we use a convention $r_B, r_f \ge 0, 0\le \delta_B, \delta_f < 2\pi$),  
\beq\label{ratios}
\frac{A(B\to D^{(*)0}K^{(*)})}{A(B\to \bar D^{(*)0}K^{(*)})} = r_Be^{i(\delta_B +\gamma)}~,
~~~~\frac{A(D^0\to f_D)}{A(\bar D^0\to f_D)} = r_fe^{i\delta_f}~.
\eeq
In three-body decays of $B$ and $D$ mesons, such as $B\to DK\pi$ and $D\to K\pi\pi$, the 
two pairs of parameters $(r_B,\delta_B)$ and $(r_f,\delta_f)$ are actually functions of two 
corresponding Dalitz variables describing the kinematics of the above three-body decays. 
The sensitivity of determining $\gamma$ depends on 
$r_B$ and $r_f$ because this determination relies on an interference of $D^0$ and 
$\bar D^0$ amplitudes. For $D$ decay modes with $r_f\sim 1$ (see discussion below) 
the sensitivity increases with the magnitude of $r_B$.

For each of the eight sub-classes of processes, $B^{+,0}\to D^{(*)}K^{(*)+,0}$,
one may study a variety of final states in neutral $D$ decays. The states $f_D$ may be 
divided into four families, distinguished qualitatively
by their parameters $(r_f,\delta_f)$ defined in Eq.~(\ref{ratios}):

\medskip
(1) $f_D=$ CP-eigenstate\cite{Gronau:1990ra,Gronau:1991dp,Gronau:1998vg} 
($K^+K^-, K_S\pi^0, {\rm etc.}); r_f =1, \delta_f =0,\pi$.

(2) $f_D=$ flavorless but non-CP state\cite{Grossman:2002aq}
($K^+K^{*-}, K^{*+}K^-, {\rm etc.}); r_f ={\cal O}(1)$.

(3) $f_D=$ flavor state\cite{Atwood:1996ci} ($K^+\pi^-, K^+\pi^-\pi^0, {\rm etc.}); r_f \sim \tan^2\theta_c$.

(4) $f_D=$ 3-body self-conjugate state\cite{Giri:2003ty} ($K_S\pi^+\pi^-); r_f,\delta_f$
vary across the Dalitz\\
$~~~~~~~~~$ plane.\\
In the first family, CP-odd states occur in Cabibbo-favored $D^0$ and $\bar D^0$ decays, 
while CP-even states occur in singly Cabibbo-suppressed decays. The second family of 
states occurs in singly Cabibbo-suppressed decays, the third family occurs in Cabibbo-favored 
$\bar D^0$ decays and in doubly Cabibbo-suppressed $D^0$ decays, while the last state is 
formally a Cabibbo-favored mode for both $D^0$ and $\bar D^0$.

\medskip
The parameters $r_B$ and $\delta_B$ in $B\to D^{(*)}K^{(*)}$ depend on whether the $B$ meson is 
charged or neutral, and may differ for $K$ vs $K^*$,\cite{Dunietz:1991yd} and for $D$ vs $D^*$,
where a neutral $D^*$ can be observed in $D^*\to D\pi^0$ or $D^*\to 
D\gamma$.\cite{Bondar:2004bi} The ratio $r_B$ involves a CKM factor 
$|V_{ub}V_{cs}/V_{cb}V_{us}|\simeq 0.4$ in both $B^+$ and $B^0$ decays, and  a 
color-suppression factor in $B^+$ decays, while in $B^0$ decays both  
$\bar b\to \bar cu \bar s$ and $\bar b \to \bar uc\bar s$ amplitudes are 
color-suppressed. A rough estimate of the color-suppression factor in these 
decays may be obtained from the color-suppression measured in corresponding 
CKM-favored decays, $B\to D\pi, D^*\pi, D\rho, D^*\rho$,  where the suppression is 
found to be in the range $0.3-0.5$.\cite{Yao:2006px}
Thus, one expects $r_B(B^0) \sim 0.4,~ r_B(B^+) = (0.3-0.5) r_B(B^0)$ in all the
processes $B^{+,0}\to D^{(*)}K^{(*)+,0}$. We note that three-body $B^+$ decays, such as
$B^+\to D^0 K^+\pi^0$, are not color-suppressed, making these processes advantageous 
by their potentially large value of $r_B$ which varies in phase space.\cite{Aleksan:2002mh,Gronau:2002mu}
 
 The above comparison of $r_B(B^+)$ and $r_B(B^0)$ may be quantified more precisely
 by expressing the four ratios $r_B(B^0)/r_B(B^+)$ in $B\to D^{(*)}K^{(*)}$ in terms of 
 reciprocal ratios of known magnitudes of  amplitudes:\cite{Gronau:2004gt} 
 \beq\label{r-ratio}
\frac{r_B(B^0\to D^{(*)}K^{(*)0})}{r_B(B^+\to D^{(*)}K^{(*)+})}\simeq 
\sqrt{\frac{\b(B^+\to \bar D^{(*)0}K^{(*)+})}{\b(B^0\to \bar D^{(*)0}K^{(*)0})}}~.
\eeq
 This follows from an approximation,
 \beq\label{A0=A+}
 A(B^0\to D^{(*)0}K^{(*)0}) \simeq A (B^+\to D^{(*)0}K^{(*)+})~,
 \eeq
 where the $B^0$ and $B^+$ processes are related to each other by replacing a spectator $d$ quark 
 by a $u$ quark. While formally Eq.~(\ref{A0=A+}) is not an isospin prediction, it may be  obtained 
using an isospin  triangle relation,\cite{Gronau:1998un}
\beq\label{iso}
A(B^0\to D^{(*)0}K^{(*)0}) = A (B^+\to D^{(*)0}K^{(*)+}) + A(B^+\to D^{(*)+}K^{(*)0}),
\eeq
and neglecting the second amplitude on the right-hand-side which is ``pure annihilation".\cite{Blok:1997yj} 
This amplitude is expected to be suppressed by a factor of four or five relative to the other two 
amplitudes appearing in (\ref{iso}) which are color-suppressed. Evidence for this kind of suppression  
is provided by corresponding ratios of CKM-favored amplitudes,\cite{Yao:2006px}
$|A(B^0\to D^-_s K^+)/\sqrt{2}A(\bar D^0\pi^0)|=0.23 \pm 0.03,
|A(B^0\to D^{*-}_s K^+)/\sqrt{2}A(\bar D^{*0}\pi^0)|<0.24$.  

Applying Eq.~(\ref{r-ratio}) to measured branching ratios,\cite{Yao:2006px,Aubert:2006qn} one 
finds 
\beq\label{ratio-r}
\frac{r_B(B^0)}{r_B(B^+)} =
\left\{ \begin{array}{cccc}
~~~B\to DK~~~& ~~B\to DK^*~~~& ~~B\to D^*K~~~ & ~~B \to D^*K^* \\
2.9 \pm 0.4 & ~~3.7 \pm 0.3~~~~& ~~> 2.2~~~&~~ >3.0 \\
\end{array} \right.
\eeq
This agrees with values of $r_B(B^0)$ near 0.4 and $r_B(B^+)$
between 0.1 and 0.2.
Note that in spite of the expected  larger values of $r_B$ in $B^0$ decays,
from the point of view of statistics alone (without considering the question of flavor 
tagging and the efficiency of detecting a $K_S$ in $B^0\to D^{(*)}K^0$), 
$B^+$ and $B^0$ decays may fare comparably when studying $\gamma$. 
This follows from (\ref{A0=A+}) because the statistical error on $\gamma$ scales roughly 
as the inverse of the smallest of the two interfering amplitudes.

We will now discuss the actual manner by which $\gamma$ can be determined 
using {\em separately} three of the above-mentioned families of final states $f_D$. 
We will mention advantages and disadvantages in each case.
For illustration of the method we will consider $B^+\to f_DK^+$.  We will also summarize
the current status of these measurements in all eight decay modes 
$B^{+,0}\to D^{(*)}K^{(*)+,0}$. 

\subsection{$f_D=$ CP-eigenstates}
One considers four observables consisting of  
two charge-averaged decay rates for even and odd CP states, normalized by the decay 
rate into a $D^0$ flavor state,
\beq
R_{{\rm CP}\pm}  \equiv  \frac{\Gamma(D_{{\rm CP}\pm} K^-) + 
\Gamma(D_{{\rm CP}\pm} K^+)}{\Gamma(D^0 K^-)}~,
\eeq
and two CP asymmetries for even and odd CP states,
\beq
A_{{\rm CP}\pm}  \equiv  \frac{\Gamma(D_{{\rm CP}\pm} K^-) - 
\Gamma(D_{{\rm CP}\pm} K^+)}{\Gamma(D_{{\rm CP}\pm} K^-) 
+ \Gamma(D_{{\rm CP}\pm} K^+)}~.
\eeq
In order to avoid dependence of $R_{{\rm CP}\pm}$ on errors in $D^0$ and $D_{\rm CP}$ 
branching ratio measurements one uses a definition of $R_{{\rm CP}\pm}$ in terms of ratios of 
$B$ decay branching ratios into $DK$ and $D\pi$ final states.\cite{Gronau:2002mu}
The four observables $R_{{\rm CP}\pm} $ and $A_{{\rm CP}\pm}$ provide three 
independent equations for $r_B, \delta_B$ and $\gamma$,
\bea
R_{{\rm CP}\pm} & = & 1 + r_B^2 \pm 2r_B\cos\delta_B\cos\gamma~,\\
A_{{\rm CP}\pm} & = & \pm 2r_B \sin\delta_B \sin\gamma/R_{{\rm CP}\pm}~.
\eea

While in principle this is the simplest and most precise method for extracting 
$\gamma$, up to a discrete ambiguity, in practice this method is sensitive to $r_B^2$, 
because $(R_{{\rm CP}+}+R_{{\rm CP}-})/2 = 1 +r_B^2$.
This becomes very difficult for charged $B$ decays where one expects 
$r_B\sim 0.1-0.2$, but may be feasible for neutral $B$ decays where $r_B\sim 0.4$.
An obvious signature for a non-zero value of $r_B$ would be observing a difference between 
$R_{{\rm CP}+}$ and $R_{{\rm CP}-}$ which is linear in this quantity. 

Studies of $B^+\to D_{\rm CP} K^+, B^+ \to D_{\rm CP}K^{*+}$ and $B^+\to 
D^*_{\rm CP}K^+$ have been carried out recently,\cite{Aubert:2005cc,Aubert:2005rw,Abe:2006hc}  
each consisting of a few tens of events. A nonzero difference 
$R_{{\rm CP}+}-R_{{\rm CP}-}$ at $2.6$ standard deviations, measured in
$B^+\to D_{\rm CP}K^{*+}$,\cite{Aubert:2005cc} is probably a statistical fluctuation.  A larger difference 
is anticipated in $B^0\to D_{CP}K^{*0}$, as the value of $r_B$ in this process
is expected to be three or four times larger than in $B^+\to DK^{*+}$. [See Eq. (\ref{ratio-r}).] 
Higher statistics is required for a measurement of $\gamma$ using this method. 

\subsection{$f_D=$ flavor state}
Consider a flavor state $f_D$ in Cabibbo-favored $\bar D^0$ decays, accessible also to
doubly Cabibbo-suppressed $D^0$ decays, such that one has $r_f\sim \tan^2\theta_c$
in Eq.~(\ref{ratios}).  One studies the ratio of two charge-averaged 
decay rates, for decays into $\bar f_DK$ and $f_DK$,
\beq
R_f  \equiv \frac{\Gamma(f_DK^-) + \Gamma(\bar f_DK^+)}
{\Gamma(\bar f_DK^-) + \Gamma(f_DK^+)}~,
\eeq 
and the CP asymmetry,
\beq
A_f  \equiv  \frac{\Gamma(f_DK^-) - \Gamma(\bar f_DK^+)}
{\Gamma(f_DK^-) + \Gamma(\bar f_DK^+)}~.
\eeq 
These observables are given by 
\bea\label{Rf}
R_f & = & r_B^2 + r^2_f + 2r_B\,r_f\,\cos(\delta_B - \delta_f)\,\cos\gamma~,\\
A_f  &= &  2r\,r_f\,\sin(\delta_B - \delta_f)\,\sin\gamma/R_f~,
\eea
where a multiplicative correction $1+{\cal O}(r_Br_f)\sim 1.01$ has been neglected
in (\ref{Rf}).

These two observables involve three unknowns, $r_B, \delta_B-\delta_f$ and $\gamma$.
One assumes $r_f$ to be given by the measured ratio of 
doubly Cabibbo-suppressed and Cabibbo-favored branching ratios.
Thus, one needs at least two flavor states, $f_D$ and $f'_D$, for which two pairs of
observables ($R_f, A_f$) and ($R_{f'}, A_{f'}$)  provide four equations for the four unknowns,
$r_B, \delta_B-\delta_f, \delta_B - \delta_{f'}, \gamma$.
The strong phase differences $\delta_f, \delta_{f'}$ can actually be measured  at a 
$\psi''$ charm factory,\cite{Silva:1999bd} thereby reducing the number of unknowns to three.

While the decay rate in the numerator of $R_f$ is rather low, the asymmetry $A_f$ may 
be large for small values of $r_B$ around 0.1, as it involves two amplitudes with a relative 
magnitude $r_f/r_B$. 

So far, only upper bounds have been measured for $R_f$ implying upper limits on $r_B$ in 
several processes, $r_B(B^+\to DK^+)<0.2$,\cite{Aubert:2005pj,Abe:2005gi,Aubert:2006ga}
$r_B(B^+\to D^*K^+)  < 0.2$,\cite{Aubert:2005pj}~$r(B^+\to 
DK^{*+})<0.4$,\cite{Aubert:2005cr} and 
$r_B(B^0\to DK^{*0}) < 0.4$.\cite{Aubert:2006qn,Krokovny:2002ua}  
Further constraints on $r_B$ in the first three processes have been obtained by studying 
D decays into CP-eigenstates and into the state $K_S\pi^+\pi^-$.
Using $r_B(B^0\to DK^{*0})/r_B(B^+\to DK^{*+}) = 3.7 \pm 0.3$ in (\ref{ratio-r}) and assuming that  
$r_B(B^+\to DK^{*+})$ is not smaller than about 0.1, one may conclude
that a nonzero measurement of $r_B(B^0\to DK^{*0})$ should be measured soon. The signature 
for $B^0\to D^0K^{*0}$ events would be two kaons with opposite charges.

\subsection{$f_D=K_S\pi^+\pi^-$}
The amplitude for $B^+ \to (K_S\pi^+\pi^-)_DK^+$ is a function of the two invariant-mass 
variables, $m^2_{\pm} \equiv (p_{K_S}+p_{\pi^{\pm}})^2$, and may be written as
\beq
A(B^+ \to (K_S\pi^+\pi^-)_DK^+) =
f(m^2_+,m^2_-) + r_Be^{i(\delta_B+\gamma)}f(m^2_-,m^2_+)~.
\eeq
In $B^-$ decay one replaces $m_+ \leftrightarrow m_-,~\gamma \to -\gamma$.
The function $f$ may be written as a sum of about twenty resonant and nonresonant 
contributions  modeled to describe the amplitude for flavor-tagged $\bar D^0\to K_S\pi^+\pi^-$
which is measured separately.\cite{Poluektov:2006ia,Aubert:2006am}
This introduces a model-dependent uncertainty in the analysis. Using the measured 
function $f$ as an input and fitting the rates for $B^{\pm}\to (K_S\pi^+\pi^-)_DK^{\pm}$  
to the parameters, $r_B,\delta_B$ and $\gamma$, one then determines these three 
parameters. 

The advantage of using $D\to K_S\pi^+\pi^-$ decays over CP and flavor states is 
being Cabibbo-favored and involving regions in phase space with a potentially large
interference between $D^0$ and $\bar D^0$ decay amplitudes.
The main disadvantage is the uncertainty introduced by modeling the function $f$.

Two recent analyses of comparable statistics by Belle and Babar, combining $B^\pm\to DK^\pm, 
B^\pm \to D^*K^\pm$ and $B^\pm \to DK^{*\pm}$, obtained values\,\cite{Poluektov:2006ia}  
$\gamma = [53^{+15}_{-18}\pm 3 \pm  9 ({\rm model})]^\circ$ and
$\gamma = [92\pm 41\pm 11\pm 12 ({\rm model})]^\circ$.\cite{Aubert:2006am} 
[This second value does not use the process $B^+\to D(K_S\pi^+)_{K^*}$,  
also studied by the same group,\cite{Aubert:2005yj}.]
The larger errors in the second analysis are correlated with smaller values of the extracted 
parameters $r_B$ in comparison with those extracted in the first study.
The model-dependent errors may be reduced by studying at CLEO-c the 
decays $D_{CP\pm}\to K_S\pi^+\pi^-$, providing further information
on strong phases in $D$ decays.\cite{Silva:1999bd}

\medskip\noindent
{\bf Conclusion}: The currently most precise value of $\gamma$ is 
$\gamma = [53^{+15}_{-18}\pm 3 \pm  9 ({\rm model})]^\circ$, obtained from
$B^{\pm}\to D^{(*)}K^{(*)\pm}$ using $D\to K_S\pi^+\pi^-$. These errors may be reduced 
in the future by combining the study of {\em all $D$ decay modes} in 
$B^{+,0}\to D^{(*)}K^{(*)+,0}$.
The decay $B^0\to DK^{*0}$ seems to carry a high potential because of its expected 
large value of $r_B$. Decays $B^0\to D^{(*)}K^0$ may also turn useful, as they have
been shown to provide information on $\gamma$ without the need for flavor tagging
 of the initial $B^0$.\cite{Gronau:2004gt,Gronau:2007bh}

\section{The currently most precise determination of $\gamma$: $B\to\pi\pi, \rho\rho, \rho\pi$}
\subsection{$B\to\pi\pi$}

The amplitude for $B^0\to\pi^+\pi^-$ contains two terms, 
conventionally denoted ``tree" ($T$) and ``penguin" ($P$) 
amplitudes,~\cite{Gronau:1989ia,London:1989ph} 
involving a weak CP-violating phase $\gamma$ and a strong CP-conserving phase 
$\delta$, respectively: 
\beq
A(\bo \to \pi^+ \pi^-) =  |T| e^{i \gamma} + |P| e^{i \delta}~.
\eeq
Time-dependent decay rates, for an initial $B^0$ or a $\ob$, are
given by
\beq\label{Asym}
\Gamma(B^0(t)/\ob(t)\to\pi^+\pi^-) = 
e^{-\Gamma t}\Gamma_{\pi^+\pi^-}\,\left [ 1\pm \cpp\cos\Delta m t \mp \spp\sin\Delta m t\right ]~,
\eeq 
where
\beq\label{SC}
\spp= \frac{2 {\rm Im}(\lpp)}{1 + |\lpp| ^2}~,~~~
\cpp = \frac{1 - |\lpp|^2}{1 + |\lpp|^2}~,~~~
\lpp \equiv  e^{-2i \beta} \frac{A(\ob \to \pi^+ \pi^-)}
{A(B^0 \to \pi^+ \pi^-)}~.
\eeq
One has\cite{Gronau:1989ia}
\bea
\spp & = & \sin 2\alpha + 2|P/T|\cos 2\alpha\sin(\beta+\alpha)\cos\delta 
+{\cal O}(|P/T|^2)~,\cr
\cpp & = & 2|P/T|\sin(\beta + \alpha)\sin\delta + {\cal O}(|P/T|^2)~.
\eea
This tells us two things:\\
(1) The deviation of $\spp$ from $\sin 2\alpha$ and 
the magnitude of $\cpp$ increase with $|P/T|$, which 
can be estimated to be $|P/T|\sim 0.5$ by comparing $B\to \pi\pi$ 
rates with penguin-dominated $B\to K\pi$ rates.\cite{Gronau:2004ej}\\ 
(2) $\Gamma_{\pi^+\pi^-}$, $\spp$ and $\cpp$ are 
insufficient for determining $|T|, |P|, \delta$ and $\gamma$ (or $\alpha$).\\
Further information on these quantities may be obtained by applying isospin 
symmetry to all $B\to\pi\pi$ decays.

In order to carry out an isospin analysis,\cite{Gronau:1990ka} one uses 
the fact that the three physical $B\to\pi\pi$ decay amplitudes and the three $\bar B\to\pi\pi$ 
decay amplitudes, depending each on two isospin amplitudes, obey triangle relations of
the form,
\beq\label{isotr}
A(B^0\to \pi^+\pi^-)/\s + A(B^0\to \pi^0\pi^0)-A(B^+\to \pi^+\pi^0)=0~~.
\eeq
Furthermore, the penguin amplitude is pure $\Delta I= 1/2$; hence the $\Delta I=3/2$ 
amplitude carries a week phase $\gamma$, $A(B^+\to\pi^+\pi^0)=e^{2i\gamma}
A(B^-\to\pi^-\pi^0)$. Defining $\sin 2\alpha_{\rm eff} \equiv S_{+-}/(1 - C^2_{+-})^{1/2}$, 
the difference $\alpha_{\rm eff}-\alpha$ is 
then determined by an angle between corresponding sides of 
the two isospin triangles sharing a common base, $|A(B^+\to\pi^+\pi^0)|=
|A(B^-\to\pi^-\pi^0)|$. A sign ambiguity in $\alpha_{\rm eff}-\alpha$ is resolved by 
two model-independent features which are confirmed experimentally, $|P|/|T|\le 1, 
|\delta|\le \pi/2$. This implies $\alpha < \alpha_{\rm eff}$.\cite{Gronau:2004sj}

\begin{table}[h]
\tbl{Branching ratios and CP asymmetries in $B\to\pi\pi,~B\to\rho\rho$.}
{\begin{tabular}{@{}cccc@{}} 
\toprule
Decay mode & Branching ratio ($10^{-6}$) & $A_{CP}=-C$ &
$S$ \\
\colrule
$B^0\to\pi^+\pi^-$ & $5.16\pm 0.22$ & $0.38 \pm 0.07$ & $-0.61 \pm 0.08$ \\
$B^+\to\pi^+\pi^0$ & $5.7\pm 0.4$ & $0.04\pm 0.05$ &  \\
$B^0\to\pi^0\pi^0$ & $1.31\pm 0.21$ &  $0.36^{+0.33}_{-0.31}$ &  \\ 
$B^0\to\rho^+\rho^-$ & $23.1^{+3.2}_{-3.3}$ & $0.11\pm 0.13$ & $-0.06\pm 0.18$ \\
$B^+\to\rho^+\rho^0$ & $18.2\pm 3.0$ & $-0.08\pm 0.13$ & \\
$B^0\to\rho^0\rho^0$ & $1.07 \pm 0.38$ & & \\
\botrule
\end{tabular}}
\end{table}

Current CP-averaged branching ratios and CP asymmetries for $B\to\pi\pi$ and 
$B\to\rho\rho$ decays are given in Table I,\cite{HFAG} where $A_{CP}\equiv-C$ 
for decays to CP eigenstates.  An impressive experimentally progress has been 
achieved in the past two years in extracting a precise value for $\alpha_{\rm eff}$,  
$\alpha_{\rm eff}=(110.6^{+3.6}_{-3.2})^\circ$.  However, the error on 
$\alpha_{\rm eff}-\alpha$ using the isospin triangles is still large. 
An upper bound, given by CP-averaged rates and a direct CP asymmetry in 
$B^0\to\pi^+\pi^-$,\cite{Gronau:2001ff,Grossman:1997jr}
\beq\label{bound}
\cos 2(\alpha_{\rm eff}-\alpha) \ge \frac{\left( {1\over 2}\Gamma_{+-} + \Gamma_{+0} - 
\Gamma_{00} \right)^2 -
 \Gamma_{+-}\Gamma_{+0}}{\Gamma_{+-} \Gamma_{+0} \sqrt{1-C^2_{+-}}}~,
\eeq
leads to $0<\alpha_{\rm eff}-\alpha<31^\circ$ at $1\sigma$. Adding in
quadrature the error in $\alpha_{\rm eff}$ and the uncertainty in $\alpha-\alpha_{\rm eff}$,
this  implies $\alpha=(95\pm 16)^\circ$ or $\gamma =(64\pm 16)^\circ$ by .
A similar central value but a smaller error, $\alpha = (97\pm 11)^\circ$, has been reported
recently by the Belle Collaboration.\cite{Ishino:2006if}
The possibility that a penguin amplitude in $B^0\to\pi^+\pi^-$ may lead to a large CP asymmetry 
S for values of $\alpha$ near $90^\circ$ where $\sin 2\alpha=0$ was anticipated fifteen years ago.\cite{Gronau:1992rm} 

The bound on $\alpha_{\rm eff}-\alpha$ may be improved considerably by measuring 
a nonzero direct CP asymmetry in $B^0\to\pi^0\pi^0$. This asymmetry can be shown
to be {\em large and positive} (see Eq.~(\ref{ACPpi0pi0}) in Sec.~5.2), implying 
a large rate for $\bar B^0$ but a small rate for $B^0$. Namely, the triangle (\ref{isotr}) is 
expected to be squashed, while the $\bar B$ triangle is roughly equal sided.

An alternative way of treating the penguin amplitude in 
$B^0\to\pi^+\pi^-$ is by combining within flavor SU(3) the decay rate and asymmetries in 
this process with rates and asymmetries in $B^0\to K^0\pi^+$ or 
$B^0\to K^+\pi^-$.\cite{Gronau:2004ej} The ratio of $\Delta S=1$ and $\Delta S=0$
tree amplitudes in these processes, excluding CKM factors, is taken to be given by 
$f_K/f_\pi$ assuming factorization, while the ratio of corresponding
penguin amplitudes is allowed to vary by $\pm 0.22$ around one.
A current update of this rather conservative analysis obtains~\cite{GR}
 \beq\label{gamma-pipi}
 \gamma = (73\pm 4^{+10}_{-8})^\circ~,
 \eeq
 where the first error is experimental, while the second one is due to an uncertainty in 
 SU(3) breaking. A discussion of SU(3) breaking factors relating $B^0\to\pi^+\pi^-$ and 
 $B^0\to K^+\pi^-$ is included in Section 5.2.

\subsection{$B\to\rho\rho$}

Angular analyses of the pions in $\rho$ decays have shown that 
$B^0\to \rho^+\rho^-$ is dominated almost $100\%$ by longitudinal polarization~\cite{HFAG}. 
This simplifies the isospin analysis of CP asymmetries in these decays to becoming similar 
to $B^0\to\pi^+\pi^-$. The advantage of $B\to \rho\rho$ over $B\to \pi\pi$ is  the 
relative small value of $\b(\rho^0\rho^0)$ in comparison with $\b(\rho^+\rho^-)$ and
$\b(\rho^+\rho^0)$ (see Table I), indicating a smaller 
$|P/T|$ in $B\to\rho^+\rho^-$ ($|P/T|< 0.3$~\cite{Charles:2006yw}) than in $B^0\to\pi^+\pi^-$
($|P/T|\sim 0.5$\,\cite{Gronau:2004ej}). Eq.~(\ref{bound}) leads to an upper bound
on $\alpha_{\rm eff}-\alpha$ in $B\to\rho\rho$, $0<\alpha_{\rm eff}-\alpha< 17^\circ$ (at $1\sigma$). 
The asymmetries for longitudinal $\rho$'s given in Table I imply 
$\alpha_{\rm eff} = (91.7^{+5.3}_{-5.2})^\circ$. Thus, one finds $\alpha = (83 \pm 10)^\circ$ 
or $\gamma = (76\pm 10)^\circ$ by adding errors in quadrature. 

A stronger bound on $|P/T|$ in $B^0\to\rho^+\rho^-$, leading to a more precise 
value of $\gamma$,  may be obtained by relating this process to $B^+\to K^{*0}\rho^+$ 
within flavor SU(3).~\cite{Beneke:2006rb}
One uses the branching ratio and fraction of longitudinal rate measured for this 
process~\cite{HFAG}, $\b(K^{*0}\rho^+)=(9.2 \pm 1.5)\times 10^{-6}, f_L(K^{*0}\rho^+)=
0.48\pm 0.08$, to normalize 
the penguin amplitude in $B^0\to\rho^+\rho^-$. Including a conservative uncertainty from 
SU(3) breaking and smaller amplitudes, one finds a value 
\beq\label{gamma-rhorho}
\gamma = (71.4^{+5.8}_{-8.8}~^{+4.7}_{-1.7})^\circ~,
\eeq
where the first error is experimental and the second one theoretical. 

The current small theoretical error in $\gamma$ requires including isospin breaking 
effects in studies based on isospin symmetry. The effect of electroweak penguin 
amplitudes on the isospin analyses of $B\to\pi\pi$ and $B\to \rho\rho$ has been calculated 
and was found to move $\gamma$ slightly higher by an amount  
$\Delta\gamma_{\rm EWP}=1.5^\circ$.\cite{Buras:1998rb,Gronau:1998fn}
Other corrections, relevant to methods using $\pi^0$ and $\rho^0$,
includng $\pi^0$-$\eta$-$\eta'$ mixing, $\rho$-$\omega$ mixing, and a small 
$I=1$ $\rho\rho$ contribution allowed by the $\rho$-width, are each smaller than one degree.\cite{Gardner:1998gz,Gronau:2005pq,Falk:2003uq} 

\medskip\noindent
{\bf Conclusion}:
Taking an average of the two values of $\gamma$ in (\ref{gamma-pipi}) and 
(\ref{gamma-rhorho}) obtained from $B^0\to\pi^+\pi^-$ and $B^0\to \rho^+\rho^-$, and
including the above-mentioned EWP correction, one finds 
\beq\label{gamma}
\gamma=(73.5\pm 5.7)^\circ~.
\eeq
A third method of measuring $\gamma$ (or $\alpha$) in time-dependent Dalitz analyses of
$B^0\to (\rho\pi)^0$ involves a much larger error,\cite{Snyder:1993mx} and has a small effect
on the overall averaged value of the weak phase.
We note that $\sin\gamma$ is close to one and its relative error is only $3\%$, the same as the 
relative error in $\sin 2\beta$ and slightly smaller than the relative error in $\sin\beta$.

\section{Rates, asymmetries, and $\gamma$ in $B\to K\pi$}
\subsection{Extracting $\gamma$ in $B\to K\pi$}

The four decays $B^0\to K^+\pi^-, B^0\to K^0\pi^0, B^+\to K^0\pi^+, B^+\to K^+\pi^0$ 
involve a potential for extracting $\gamma$, 
provided that one is sensitive to interference between a dominant isoscalar penguin 
amplitude and a small tree amplitude contributing to these processes.
This idea has led to numerous suggestions for determining $\gamma$ in these decays
starting with a proposal made in 1994.\cite{Gronau:1994rj,Gronau:1994bn}  
An interference between penguin and  tree amplitudes may be identified in two ways:

(1) Two different properly normalized $B\to K\pi$ rates.

(2)  Nonzero direct CP asymmetries.

\begin{table}[h]
\tbl{Branching ratios and asymmetries in $B\to K\pi$.}
{\begin{tabular}{@{}ccc@{}} 
\toprule
Decay mode & Branching ratio ($10^{-6}$) & $A_{CP}$ \\
\colrule
$B^0\to K^+\pi^-$ & $19.4\pm 0.6$ & $-0.097 \pm 0.012$  \\
$B^+\to K^+\pi^0$ & $12.8\pm 0.6$ & $0.047\pm 0.026$  \\
$B^+\to K^0\pi^+$ & $23.1\pm 1.0$ &  $0.009\pm 0.025$  \\ 
$B^0\to K^0\pi^0$ & $10.0\pm 0.6$ & $-0.12\pm 0.11$ \\
\botrule
\end{tabular}}
\end{table}

\medskip\noindent
Current branching ratios and CP asymmetries are summarized in Table II.\cite{HFAG}
Three ratios of rates, calculated using the ratio of $B^+$ and $B^0$ lifetimes,
$\tau_+/\tau_0 = 1.076 \pm 0.008$,\cite{HFAG} are:
\bea
R & \equiv & \frac{\Gamma(B^0 \to K^+ \pi^-)}
{\Gamma(B^+ \to K^0 \pi^+)} = 0.90 \pm 0.05~,\cr
R_c  & \equiv  & \frac{2\Gamma(B^+ \to K^+ \pi^0)}
{\Gamma(B^+ \to K^0 \pi^+)} = 1.11 \pm 0.07~,\cr
R_n  & \equiv & \frac{\Gamma(B^0 \to K^+ \pi^-)}
{2\Gamma(B^0 \to K^0 \pi^0)} = 0.97 \pm 0.07~.
\eea
The largest deviation from one, observed in the ratio $R$ at 2$\sigma$,
is insufficient for claiming unambiguous evidence for  a non-penguin contribution.
An upper limit, $R< 0.965$ at $90\%$ confidence level, would imply  
$\gamma \le 79^\circ$ using $\sin^2\gamma \le R$,\cite{Fleischer:1997um} which 
neglects however ``color-suppressed" EWP contributions.\cite{Gronau:1997an}
As we will argue now, these contributions and ``color-suppressed" tree amplitudes
are actually not suppressed as naively expected.

The nonzero asymmetry measured in $B^0\to K^+\pi^-$ provides first evidence for 
an interference between penguin ($P'$) and tree ($T'$) amplitudes with
a nonzero relative strong phase. Such an interference occurs also in $B^+\to K^+\pi^0$
where no asymmetry has been observed. An assumption that other contributions to the latter 
asymmetry are negligible has raised some questions about the validity of the CKM 
framework. In fact,  a color-suppressed  tree amplitude ($C'$), also occurring in 
$B^+\to K^+\pi^0$,\cite{Gronau:1994rj} resolves this ``puzzle" if this amplitude is 
comparable in magnitude to $T'$. Indeed, several studies have shown that this is the 
case,\cite{Chiang:2004nm,Baek:2004rp,Buras:2003dj,Li:2005kt,Beneke:2005vv}
also implying that color-suppressed and color-favored EWP amplitudes are of comparable magnitudes.\cite{Gronau:1998fn}
For consistency between the two CP asymmetries in $B^0\to K^+\pi^-$ and 
$B^+\to K^+\pi^0$, the strong phase difference between $C'$ and $T'$ must be 
negative and cannot be very small.\cite{Gronau:2006ha} This seems to stand in 
contrast to QCD calculations using a factorization 
theorem.\cite{Beneke:1999br,Bauer:2004tj,Beneke:2005vv}

The small asymmetry $A_{CP}(B^+\to K^+\pi^0)$ implies bounds on the sine of the strong
phase difference $\delta_c$ between $T'+C'$ and $P'$. The cosine of this phase
affects $R_c-1$ involving the decay rates for $B^+\to K^0\pi^+$ and
$B^+\to K^0\pi^+$. A question studied recently is whether the
two upper bounds on $|\sin\delta_c|$ and $|\cos\delta_c|$ are consistent with 
each other or, perhaps, indicate effects of NP. Consistency was shown by 
proving a sum rule involving $A_{CP}(B^+\to K^+\pi^0)$
and $R_c-1$, in which an electroweak penguin (EWP) amplitude plays an important role.  
We will now present a proof of the sum rule, which may provide important information
on $\gamma$.\cite{Gronau:2006ha} 

The two amplitudes for $B^+\to K^0\pi^+, K^+\pi^0$ are given in terms of topological 
contributions including $P', T'$ and $C'$,
\bea\label{AmpKpi}
A(B^+\to K^0\pi^+) & = & (P'-\frac{1}{3}P'^c_{EW}) + A'~,\cr
A(B^+\to K^+\pi^0) & = & (P'-\frac{1}{3}P'^c_{EW}) + (T'+P'^c_{EW}) +(C'+P'_{EW}) +A'~,
\eea
where $P'_{EW}$ and $P'^c_{EW}$ are color-favored and color-suppressed EWP 
contributions. The small annihilation amplitude $A'$ and a small $u$ quark contribution 
to $P'$ involving a CKM factor $V^*_{ub}V_{us}$ will be neglected ($|V^*_{ub}V_{us}|/|V^*_{cb}V_{cs}|=0.02$). Evidence for the smallness of these terms can be found in the 
small CP asymmetry measured for $B^+\to K^0\pi^+$. Large terms would require 
rescattering and a sizable strong phase difference between these terms and $P'$.

Flavor SU(3) symmetry relates $\Delta I=1, I(K\pi)=3/2$ electroweak  penguin and tree 
amplitudes through a calculable ratio $\delta_{EW}$~\cite{Gronau:1998fn,Neubert:1998pt},
\bea\label{eqn:delta_EW}
T' + C' + P'_{EW} + P'^c_{EW} & = & (T' + C')(1-\delta_{EW}e^{-i\gamma})~~,
\nonumber\\
\delta_{EW} & = & -\frac{3}{2}\frac{c_9 + c_{10}}{c_1 + c_2}
\frac{|V^*_{tb}V_{ts}|}{|V^*_{ub}V_{us}|} = 0.60 \pm 0.05~~.
\eea
The error in $\delta_{EW}$ is dominated by the current uncertainty in 
$|V_{ub}|/|V_{cb}| = 0.104 \pm 0.007$~\cite{Yao:2006px}, including also a 
smaller error from SU(3) breaking estimated using QCD factorization.
Eqs.~(\ref{AmpKpi}) and (\ref{eqn:delta_EW}) imply
\cite{Gronau:2001cj}
\beq \label{eqn:Rc}
R_c = 1 - 2 r_c \cos \delta_c (\cos \gamma - \delta_{\rm EW})
+ r_c^2(1 - 2 \delta_{\rm EW} \cos \gamma + \delta_{\rm EW}^2)~,~~
\eeq
\beq \label{eqn:Acp}
A_{CP}(B^+ \to K^+ \pi^0) =  - 2 r_c \sin \delta_c \sin \gamma /R_c~~,
\eeq
where $r_c\equiv |T'+C'|/|P'-\frac{1}{3}P'^c_{EW}|$ and $\delta_c$ is the strong phase difference
between $T'+C'$ and $P'-\frac{1}{3}P'^c_{EW}$.  

The parameter $r_c$ is calculable in terms of measured
decay rates, using broken flavor SU(3) which relates $T'+C'$ and
$T+C$ dominating $B^+\to \pi^+\pi^0$ by a factorization factor $f_K/f_\pi$
(neglecting a tiny EWP term in $B^+\to \pi^+\pi^0$),\cite{Gronau:1994bn}
\beq\label{T'+C'}
|T'+C'| = \sqrt{2}\frac{V_{us}}{V_{ud}}\frac{f_K}{f_\pi}|A(B^+\to \pi^+\pi^0)|~~.
\eeq
Using branching ratios from Tables I and II, one finds
\beq 
r_c = \sqrt{2}\frac{V_{us}}{V_{ud}}\frac{f_K}{f_\pi}
\sqrt{\frac{\b(B^+\to\pi^+\pi^0)}
{\b(B^+\to K^0\pi^+)}} = 0.198 \pm 0.008~~.
\eeq
The error in $r_c$ does not include an uncertainty from assuming 
factorization for SU(3) breaking in $T'+C'$.
While this assumption should hold well for $T'$, it may not be a good
approximation for $C'$ which as we have mentioned is comparable in 
magnitude to $T'$ and carries a strong phase relative to it.
Thus one should allow a $10\%$ theoretical error when using factorization 
for relating $B \to K \pi$ and $B \to \pi \pi$ $T+C$ amplitudes, so that
\beq
r_c =0.20 \pm 0.01~({\rm exp}) \pm 0.02~({\rm th})~~.
\eeq 

Eliminating $\delta_c$ in Eqs.~(\ref{eqn:Rc}) and (\ref{eqn:Acp}) by retaining
terms which are linear in $r_c$, one finds
\beq \label{eqn:sr}
\left( \frac{R_c-1}{\cos \gamma - \delta_{\rm EW}} \right)^2 +
\left( \frac{A_{CP}(B^+ \to K^+ \pi^0)}{\sin \gamma} \right)^2 = (2r_c)^2 +
{\cal O}(r_c^3)~~.
\eeq
This sum rule implies that at least one of the two terms whose
squares occur on the left-hand-side must be sizable, of the order of
$2r_c=0.4$. The second term, $|A_{CP}(B^+\to K^+\pi ^0)|/\sin\gamma$, is
already smaller than $\simeq 0.1$, using the current $2\sigma$ bounds on $\gamma$ and  
$|A_{CP}(B^+\to K^+\pi^0)|$. Thus, the first term must provide a
dominant contribution. For $R_c\simeq 1$, this implies $\gamma\simeq 
\arccos\delta_{EW} \simeq (53.1\pm 3.5)^\circ$. This range is expanded by including 
errors in $R_c$ and $A_{CP}(B^+\to K^+\pi^0)$. 
For instance, an upper bound $R_c <  1.1$ would imply an inportant upper limit, 
$\gamma < 70^\circ$. Currently one only obtains  
an upper limit $\gamma \le 88^\circ$ at 
$90\%$ confidence level.\cite{Gronau:2006ha} This bound is consistent with 
the value obtained in (\ref{gamma}) from $B\to\pi\pi$ and $B\to\rho\rho$, but is not 
competitive with the latter precision. 

\medskip\noindent
{\bf Conclusion}:
The current constraint obtained from $R_c$ and $A_{CP}(B^+\to K^+\pi^0)$ is
$\gamma \le 88^\circ$ at $90\%$ confidence level. 
Further improvement in the measurement of $R_c$ (which may, in fact, be very close 
to one) is required in order to achieve a precision in $\gamma$ comparable to that 
obtained in $B\to\pi\pi, \rho\rho$. (A conclusion concerning the different CP asymmetries 
measured in $B^0\to K^+\pi^-$ and $B ^+\to K^+\pi^0$ will be given at the end of 
the next subsection.)

\subsection{Symmetry relations for $B\to K\pi$ rates and asymmetries}

The following two features imply rather precise sum rules in the CKM framework,
 both for $B\to K\pi$ decay rates and CP asymmetries:
 
(1) The dominant penguin amplitude is $\Delta I=0$.

(2) The four decay amplitudes obey a linear isospin relation,\cite{Nir:1991cu}
\beq
A(K^+\pi^-) - A(K^0\pi^+) - \s A(K^+\pi^0) + \s A(K^0\pi^0)~.
\eeq
An  immediate consequence of these features are two isospin sum rules, 
which hold up to terms which are quadratic in small ratios of
non-penguin to penguin amplitudes,\cite{Gronau:1998ep,Atwood:1997iw,Gronau:2005gz}

\bea\label{RSR}
\Gamma(K^+\pi^-) + \Gamma(K^0\pi^+) &=&
2\Gamma(K^+\pi^0) + 2\Gamma(K^0\pi^0)~,\\
\label{DSR}
\Delta(K^+\pi^-) +\Delta(K^0\pi^+) &=&
2\Delta(K^+\pi^0) + 2\Delta(K^0\pi^0)~,
\eea
where
\beq\label{Delta}
\Delta(K\pi)\equiv \Gamma(\bar B\to\bar K\bar\pi) - \Gamma(B\to K \pi)~.
\eeq

Quadratic corrections to (\ref{RSR}) have been calculated in the SM
and were found to be a few percent.\cite{Gronau:2003kj,Beneke:2003zv,Bauer:2005kd}
This is the level expected in general for isospin-breaking corrections which must therefore 
also be considered. The above two features  
imply that these $\Delta I=1$ corrections are suppressed by a small ratio of
non-penguin to penguin amplitudes and are therefore negligible.\cite{Gronau:2006eb} 
Indeed, this sum rule holds experimentally within a $5\%$ error.\cite{Gronau:2006xu}
One expects the other sum rule (\ref{DSR}) to hold at a similar precision.

The CP rate asymmetry sum rule (\ref{DSR}), relating the four CP asymmetries, 
leads to a prediction for the asymmetry in $B^0\to K^0\pi^0$ in terms of the other 
three asymmetries which have been measured with higher precision,
\beq\label{ACPK0pi0}
A_{CP}(B^0\to K^0\pi^0) = -0.140\pm 0.043~.
\eeq
While this value is consistent with experiment (see Table II),  higher accuracy in this 
asymmetry measurement is required for testing this straightforward prediction.

Relations between CP asymmetries in $B\to K\pi$ and $B\to\pi\pi$ following from
approximate flavor SU(3) symmetry of QCD~\cite{Zeppenfeld:1980ex} are not 
expected to hold as precisely as isospin relations, but may still be interesting and useful. 
An important question relevant to such relations is how to include SU(3)-breaking 
effects, which are expected to be at a level of 20-30$\%$. Here we wish to discuss 
two SU(3) relations proposed twelve 
years ago,\cite{Deshpande:1994ii,Gronau:1995qd} one of which holds experimentally 
within expectation, providing some lesson about SU(3) breaking, while the other has 
a an interesting implication for future applications of the isospin analysis in $B\to \pi\pi$. 

A most convenient proof of
SU(3) relations is based on using a diagramatic approach, in which diagrams
with given flavor topologies replace reduced SU(3) matrix elements.\cite{Gronau:1994rj}
In this language, the amplitudes for $B^0$ decays into pairs of charged or neutral pions, 
and pairs of charged or neutral  $\pi$ and $K$, are given by:
\bea
-A(B^0\to\pi^+\pi^-) & = & T+  \left (P+2P^c_{EW}/3\right ) + E + PA~,\cr
-\s A(B^0\to\pi^0\pi^0) & = & C -  \left(P-P_{EW}-P^c_{EW}/3\right)  - E - PA~,\cr
-A(B^0\to K^+\pi^-) & = & T'+ \left(P'+2P'^c_{EW}/3\right)~,\cr
-\s A(B^0\to K^0\pi^0) & = & C' - \left(P'-P'_{EW}-P'^c_{EW}/3\right)~.
\eea
The combination $E+PA$, representing exchange and penguin annihilation 
topologies, is expected to be $1/m_b$-suppressed relative to $T$ 
and $C$,\cite{Bauer:2004tj,Blok:1997yj} as demonstrated by the small branching ratio
measured for $B^0\to K^+K^-$.\cite{HFAG} This term will be neglected.

Expressing topological amplitudes in terms of CKM factors, SU(3)-invariant 
amplitudes and SU(3) invariant strong phases, one may write 
\bea
T & \equiv & V^*_{ub}V_{ud}|{\cal T}+{\cal P}_{uc}|~,~~~~~~P+2P^c_{EW}/3 \equiv 
V^*_{tb}V_{td}|{\cal P}_{tc}|e^{i\delta}~,\cr
T' & \equiv &  V^*_{ub}V_{us}|{\cal T}+{\cal P}_{uc}|~,~~~~~~P'+2P'^c_{EW}/3 \equiv
V^*_{tb}V_{ts}|{\cal P}_{tc}|e^{i\delta}~,\\
C & \equiv &  V^*_{ub}V_{ud}|{\cal C}-{\cal P}_{uc}|~,~~~~~~P-P_{EW}-P^c_{EW}/3 \equiv
V^*_{tb}V_{td}|\tilde{\cal P}_{tc}|e^{i\tilde\delta}~,\cr
C' & \equiv &  V^*_{ub}V_{us}|{\cal C}-{\cal P}_{uc}|~,~~~~~~P'-P'_{EW}-P'^c_{EW}/3 \equiv
V^*_{tb}V_{ts}|\tilde{\cal P}_{tc}|e^{i\tilde\delta}~.\nonumber
\eea
Unitarity of the CKM matrix, $V^*_{cb}V_{cd(s)} = - V^*_{tb}V_{td(s)} - V^*_{ub}V_{ud(s)}$,
has been used to absorb in $T^{(')}$ and $C^{(')}$ a penguin term 
${\cal P}_{uc}\equiv {\cal P}_u-{\cal P}_c$ multiplying $V^*_{ub}V_{ud(s)}$, while 
${\cal P}_{tc}\equiv {\cal P}_t-{\cal P}_c$ and $\tilde{\cal P}_{tc}\equiv \tilde{\cal P}_t-
\tilde{\cal P}_c$ contain two distinct combinations of EWP contributions.  Using the identity
\beq
{\rm Im}\left(V^*_{ub}V_{ud}V_{tb}V^*_{td}\right) = 
-{\rm Im}\left(V^*_{ub}V_{us}V_{tb}V^*_{ts}\right)~,
\eeq 
one finds\cite{Deshpande:1994ii,Gronau:1995qd} 
\bea\label{Delta+-}
\Delta(B^0\to K^+\pi^-) & = & -\Delta(B^0\to\pi^+\pi^-)~\\
\label{Delta00}
\Delta(B^0\to K^0\pi^0) & = & - \Delta(B^0\to\pi^0\pi^0)~,
\eea
where $\Delta$ is the CP rate difference defined in (\ref{Delta}).

Quoting products of branching ratios and asymmetries from Tables I and II, 
Eq.~(\ref{Delta+-}) reads
\beq
-1.88 \pm 0.24  = -1.96 \pm 0.37~.
\eeq
This SU(3) relation works well and requires no SU(3)-breaking. An SU(3) breaking factor 
$f_K/f_\pi$ in ${\cal T}$ but not in ${\cal P}$, or in both ${\cal T}$ and ${\cal P}$, are 
currently excluded at a level of $1.0\sigma$, or $1.75\sigma$. More precise CP asymmetry measurements in $B^0\to K^+\pi^-$ and $B^0\to \pi^+\pi^-$ are required for determining 
the pattern of SU(3) breaking in tree and penguin amplitudes. 

Using the prediction (\ref{ACPK0pi0}) of the 
$B\to K\pi$ asymmetry sum rule, Eq.~(\ref{Delta00}) predicts
\beq\label{ACPpi0pi0}
A_{CP}(B^0\to \pi^0\pi^0) = 1.07 \pm 0.38~.
\eeq
The error is dominated by current errors in CP asymmetries for $B^+\to K^0\pi^+$ and 
$B^+\to K^+\pi^0$, and to a less extent by the error in $\b(\pi^0\pi^0).$
SU(3) breaking in amplitudes could modify this prediction by a factor $f_\pi/f_K$ if this factor
applies to ${\cal C}$, and less likely by $(f_\pi/f_K)^2$. A large positive CP asymmetry, favored 
in all three cases, will affect future applications of the isospin analysis in $B\to\pi\pi$. It 
implies that while the $\bar B$ isospin triangle is roughly equal-sided, the $B$ triangle is squashed.
A twofold ambiguity in the value of $\gamma$ disappears in the limit of a flat $B$ 
triangle.\cite{Gronau:1990ra}

\medskip\noindent
{\bf Conclusion}: The isospin sum rule for $B\to K\pi$ decay rates holds well, while the
CP asymmetry sum rule predicts $A_{CP}(B^0\to K^0\pi^0)=-0.140\pm 0.043$. The different 
asymmetries in $B^0\to K^+\pi^-$ and $B^+\to K^+\pi^0$ can be explained by an  
amplitude $C'$ comparable to $T'$ and involving a relative negative strong
phase, and should not be considered a ``puzzle". 
An SU(3) relation for $B^0\to \pi\pi$ and $B^0\to K\pi$ CP asymmetries works well for charged 
modes. The corresponding relation for neutral modes  predicts a large positive asymmetry in $B^0\to\pi^0\pi^0$. Improving asymmetry measurements can provide tests for SU(3) breaking factors.

\section{Tests for small New Physics effects}
\subsection{Values of $\gamma$}

We have described three ways for extracting a value for $\gamma$ relying on  
interference of distinct pairs of quark amplitudes, $(b\to c\bar us,b\to u\bar c s),
(b\to c\bar c s, b\to u\bar u s)$ and $(b\to c\bar c d, b\to u\bar u d)$. The three pairs provide
a specific pattern for CP violation in the CKM framework, which is expected to be violated 
in many extensions of the SM. The rather precise value of $\gamma$ (\ref{gamma}) extracted from 
$B\to \pi\pi, \rho\rho, \rho\pi$ is consistent with constraints on $\gamma$ from CP conserving
measurements related to the sides of the unitarity triangle.\cite{Charles:2006yw,Bona:2006ah}
The values of $\gamma$ obtained in $B\to D^{(*)}K^{(*)}$ and $B\to K\pi$ are
consistent with those extracted in $B\to \pi\pi, \rho\rho, \rho\pi$, but are not yet 
sufficiently precise for testing  
small NP effects in charmless $B$ decays. Further experimental improvements
are required, in particular in the former two types of processes. 

While the value of $\gamma$ in $B\to D^{(*)}K^{*)}$ is not expected to be
affected by NP, the other two classes of processes involving penguin loops
are susceptible to such effects. The extraction of $\gamma$ in $B\to\pi\pi~\rho\rho$ 
assumes that $\gamma$ is the phase of a $\Delta I=3/2$ tree amplitude, while an additional
$\Delta I=3/2$ EWP contribution is included using isospin. The extracted value could 
be modified by a new $\Delta I=3/2$ effective operator originating in physics beyond the SM,
but not by a new $\Delta I=1/2$ operator.  Similarly,  the value of $\gamma$ extracted in 
$B\to K\pi$ is affected by a potential new $\Delta I=1$ operator, but not by a new
$\Delta I=0$ operator, because the amplitude 
(\ref{eqn:delta_EW}), playing an essential role in this method, is pure $\Delta I=1$. 

\subsection{$B\to K\pi$ sum rule}

Charmless $|\Delta S|=1$ $B$ and $B_s$ decays are particularly 
sensitive to NP effects, as new heavy particles at the TeV mass range may replace
the the $W$ boson and top-quark in the penguin loop dominating these 
amplitudes.\cite{Gronau:1996rv} The sum rule (\ref{RSR}) for $B\to K\pi$ decay rates provides 
a test for such effects. However, as we have argued from isospin considerations, it is 
only affected by quadratic $\Delta I=1$ amplitudes including NP contributions. Small 
NP amplitudes, contributing quadratically to the sum rule, cannot be separated 
from SM corrections, which are by themselves at a level of a few percent. This is the level
to which the sum rule has already been tested. 
We will argue below for evidence showing that potential NP 
contributions to $|\Delta S|=1$ charmless decays must be suppressed by roughly
an order of magnitude relative to the dominant $b\to s$ penguin amplitudes. 

\subsection{Values of $S,C$ in $|\Delta S|=1$ $B^0\to f_{CP}$ decays}

A class of $b\to s$ penguin-dominated $B^0$ decays to CP-eigenstates has recently 
attracted considerable attention.  This includes 
final states $XK_S$ and $XK_L$, where $X=\phi,  \pi^0, \eta',  \omega, f_0, \rho^0, K^+K^-, K_SK_S, 
\pi^0\pi^0$, for which measured asymmetries $-\eta_{CP}S$ and $C$ are quoted in Table III. 
[The asymmetries $S$ and $C=-A_{CP}$ were defined in (\ref{Asym}) for $B^0\to\pi^+\pi^-$.
Observed modes with $K_L$ in the final states obey $\eta_{CP}(XK_L)=-\eta_{CP}(XK_S)$.]
\begin{table}[h]
\tbl{Asymmetries $S$ and $C$ in $B^0\to XK_S$.}
{\begin{tabular}{@{}cccccc@{}} 
\toprule
$X$&$\phi$&$\pi^0$&$\eta'$&$\omega$&$f_0(980)$\\
$-\eta_{CP}S$&$0.39\pm 0.18$&$0.33\pm 0.21$&$0.61\pm 0.07$&
$0.48\pm 0.24$&$0.42\pm 0.17$\\
$C$&$0.01\pm 0.13$&$0.12\pm 0.11$&$-0.09\pm 0.06$&
$-0.21\pm 0.19$&$-0.02\pm 0.13$\\
\colrule
$X$&$\rho^0$&$K^+K^-$&$K_SK_S$&$\pi^0\pi^0$\\
$-\eta_{CP}S$&$0.20\pm 0.57$&$0.58^{+0.18}_{-0.13}$&
$0.58\pm 0.20$&$-0.72\pm 0.71$\\
$C$&$0.64\pm 0.46$&$0.15\pm 0.09$&$-0.14\pm 0.15$&$0.23\pm 0.54$\\
\botrule
\end{tabular}}
\end{table}
In these processes, a value $S= -\eta_{CP}\sin2 \beta$ (for states with 
CP-eigenvalue $\eta_{CP}$) is expected approximately.\cite{London:1989ph,Grossman:1996ke} 
These predictions involve hadronic uncertainties at a level of several percent, of order  
$\lambda^2,~\lambda \sim 0.2$. It has been pointed out some 
time ago\cite{Atwood:1997zr} that  it is difficult to separate these hadronic uncertainties 
within the SM from NP contributions to decay amplitudes if the latter are small. 
In the next subsection we will discuss indirect experimental evidence showing that NP 
contributions to $S$ and $C$ must be small.
Corrections to $S= -\eta_{CP}\sin2 \beta$ and values for the 
asymmetries $C$ have been calculated in the SM using methods based on QCD 
factorization\cite{Beneke:2005pu,Cheng:2005bg} 
and flavor SU(3),\cite{Chiang:2004nm,Grossman:2003qp,Gronau:2003ep}  and were found 
to be between a few percent up to above ten percent within hadronic uncertainties. 

Whereas the deviation of $S$ from $-\eta_{CP}\sin 2\beta$ is process-dependent, a generic result 
has been proven a long time ago for both $S$ and $C$, to first order in $|c/p|$,\cite{Gronau:1989ia} 
\bea\label{DeltaS}
\Delta S\equiv -\eta_{CP}S - \sin 2 \beta & = & 2\frac{|c|}{|p|}\cos 2\beta\sin\gamma\cos\Delta~,\cr
C & = & 2\frac{|c|}{|p|}\sin\gamma\sin\Delta~.
\eea
Here $p$ and $c$ are penguin and color-suppressed tree amplitudes involving a small ratio and
relative weak and strong phases $\gamma$ and $\Delta$, respectively. This implies 
$\Delta S > 0$ for $|\Delta|<\pi/2$, which can be argued for several of the above decays 
using QCD arguments\cite{Beneke:2005pu,Cheng:2005bg} or SU(3) fits.\cite{Gronau:2003ep} 
(Note that while $|p|$ is measurable in certain decay rates up to first order corrections, $|c|$ and 
$\Delta$ involve sizable hadronic uncertainties in QCD calculations.) 
In contrast to this expectation, the central values measured for $\Delta S$ are negative
for all decays. (See Table III.) Consequently, one finds an averaged value $\sin 
2\beta_{\rm eff}=0.53\pm 0.05$,\cite{HFAG} to be compared with $\sin 2\beta = 0.678\pm 0.025$.
Two measurements which seem particularly interesting are 
$-\eta_{CP}S_{\phi K_S}=0.39\pm 0.18$, where a positive correction of a few percent to 
$\sin 2\beta$  is expected in the SM,\cite{Beneke:2005pu,Cheng:2005bg} and  
$-\eta_{CP}S_{\pi^0 K_S}=0.33\pm 0.21$, where a rather large positive correction to 
$\sin 2\beta$ is expected shifting this asymmetry to a value just above $0.8$.\cite{Chiang:2004nm}

While the current averaged value of $\sin 2\beta_{\rm eff}$ is tantalizing, experimental 
errors in $S$ and $C$ must be reduced further to make a clear
case for physics beyond the SM. Assuming that the discrepancy between improved 
measurements and calculated values of $S$ and $C$ persists beyond theoretical 
uncertainties, can this provide a clue to
the underlying New Physics? Since many models could give rise to a 
discrepancy,\cite{Gronau:1996rv,Grossman:1996ke,Ciuchini:1997zp} one would
seek signatures characterizing classes of models rather than studying the effects in 
specific models. One way of classifying extensions of the SM is by the isospin 
behavior of the new effective operators contributing to $b\to s q\bar q$ transitions.

\subsection{Diagnosis of $\Delta I$ for New Physics operators} 

Four-quark operators in the effective Hamiltonian associated with NP in 
$b \to s q \bar q$ transitions can be either isoscalar or isovector operators. 
We will now discuss a study proposed recently in order to isolate 
$\Delta I=0$ or $\Delta I=1$ operators, thus determining corresponding NP 
amplitudes and CP violating phases.\cite{Gronau:2007ut} We will show that
since $S$ and $C$ in the above processes combine $\Delta I=0$ or $\Delta I=1$
contributions, separating these contributions  requires using also information from  
other two asymmetries, which are provided by  isospin-reflected decay processes.

Two $|\Delta S|=1$ charmless $B$ (or $B_s$) decay processes, related by isospin reflection,
$R_I:  u\leftrightarrow d,~\bar u\leftrightarrow -\bar d$, can always be expressed in term 
of common $\Delta I=0$ and $\Delta I=1$ amplitudes  $B$ and $A$ in the form:
\beq\label{B+-A}
A(B^+\to f) = B + A~,~~~~A(B^0\to R_If)= \pm(B - A)~.
\eeq  
A proof of this relation uses a sign change of Clebsch-Gordan coefficients under 
$m \leftrightarrow -m$.\cite{Gronau:2007ut}  The description (\ref{B+-A}) applies, 
in particular, to pairs of processes 
involving all the $B^0$ decay modes listed in Table III, and $B^+$ decay modes where 
final states are obtained by isospin reflection from corresponding $B^0$ decay modes. 
Decay rates for pairs of isospin-reflected $B$ decay processes,  and for $\bar B$ decays to 
corresponding charge conjugate final states are therefore given by (we omit inessential 
common kinematic factors),
\bea
\Gamma_+ & = & |B+A|^2~,~~~~~~~\Gamma_0 = |B-A|^2~,\cr
\Gamma_- & = & |\bar B + \bar A|^2~,~~~~~~~
\Gamma_{\bar 0} = |\bar B - \bar A|^2~.
\eea
The amplitudes $\bar B$ and $\bar A$ are related to $B$ and
$A$ by a change in sign of all weak phases, whereas strong phases are left
unchanged.  

For each pair of processes one defines four asymmetries: an isospin-dependent 
CP-conserving asymmetry,  
\beq\label{A_I}
A_I \equiv \frac{\Gamma_+ + \Gamma_- - \Gamma_0 - \Gamma_{\bar 0}}
 {\Gamma_+ + \Gamma_- + \Gamma_0 + \Gamma_{\bar 0}}~,
 \eeq
two CP-violating asymmetries for $B^+$ and $B^0$, 
\beq\label{CP-asym}
A^+_{CP}\equiv\frac{\Gamma_- - \Gamma_+}{\Gamma_-+\Gamma_+}~~,~~~~~
-C\equiv A^0_{CP}\equiv\frac{\Gamma_{\bar 0}-\Gamma_0}{\Gamma_{\bar 0} + \Gamma_0}~,
\eeq
and the time-dependent asymmetry $S$ in $B^0$ decays,
\beq \label{eqn:S}
S = \frac{2 {\rm Im} \lambda}{1 + |\lambda|^2}~~,~~~\lambda \equiv \eta_{CP}
\frac{\bar B - \bar A}{B - A} e^{- 2 i \beta}~~~,
\eeq

In the Standard Model, the isoscalar amplitude $B$ contains a dominant penguin 
contribution, $B_P$, with a CKM factor  $V^*_{cb}V_{cs}$.
The residual isoscalar amplitude,
\beq
\Delta B\equiv B-B_P~~,
\eeq
and the amplitude $A$, consist each of contributions  smaller than $B_P$ by about an 
order of 
magnitude.\cite{Beneke:1999br,Keum:2000ph,Bauer:2004tj,Ciuchini:1997hb,Gronau:1994rj}
These contributions include terms with a much smaller CKM factor 
$V^*_{ub}V_{us}$, and a higher order electroweak penguin amplitude with CKM factor 
$V^*_{tb}V_{ts}$. Thus, one expects 
\beq\label{hierarchy}
|\Delta B|  \ll  |B_P|~~,~~~~|A|\ll |B_P|~~.
\eeq 
Consequently, the asymmetries $A_I$, $A^{+,0}_{CP}$ and $\Delta S$ are expected to 
be small, of order $2|A|/|B|$ and $2|\Delta B|/|B_P|$.  In contrast,
potentially large contributions to $\Delta B$ and $A$ from NP,
comparable to $B_P$, would most likely lead to large asymmetries of order one.
An unlikely exception is the case when both $\Delta B/B_P$ and $A/B_P$ are
purely imaginary, or almost purely imaginary.   This would require very 
special circumstances such as fine-tuning in specific models.  
Excluding cancellations between NP and SM contributions in both CP-conserving 
and CP violating asymmetries, tests for the hierarchy (\ref{hierarchy}) become  
tests for the smallness of corresponding potential NP contributions to $B$ and $A$.    

There exists ample experimental information showing that asymmetries $A^+_{CP}$
are small in processes related by isospin reflection to the decay modes in Table III.
Upper limits on the magnitudes of most asymmetries are at a level 
of ten or fifteen percent [e.g., $A^+_{CP}(K^+\phi)=0.034\pm 0.044$, 
$A^+_{CP}(K^+\eta')=0.031\pm 0.026$], while others may be as large as
twenty or thirty percent [$A^+_{CP}(K^+\rho^0)=0.31^{+0.11}_{-0.10}$].  
Similar values have been measured for isospin asymmetries $A_I$ [e.g., 
$A_I(K^+\phi)=-0.037\pm 0.077$, $A_I(K^+\eta')=-0.001\pm 0.033$, 
$A_I(K^+\rho^0)=-0.16\pm 0.10$].\cite{Gronau:2007ut} Since these two types 
of asymmetries are of order $2|\Delta B|/|B_P|$ and $2|A|/|B_P|$, this confirms 
the hierarchy (\ref{hierarchy}), which can be assumed to hold also
in the presence of NP. 
 
We will take by convention the dominant penguin amplitude $B_P$ to have 
a zero weak phase and a zero strong phase, referring all other strong phases 
to it.  Writing
\beq\label{Bconvention}
B = B_P + \Delta B~~,~~~\bar B = B_P + \Delta \bar B~~,
\eeq
and expanding the four asymmetries to leading order in $\Delta B/B_P$ or $A/B_P$,
one has
\bea \label{eqn:obs1}
\Delta S & = & \cos 2 \beta\left [\frac{{\rm Im}(\bar A - A)}
{B_P} - \frac{{\rm Im}(\Delta \bar B - \Delta B)}{B_P}\right ]~~,\\
\label{eqn:obs2}
A_I   &=& \frac{{\rm Re}(\bar A + A)}{B_P}~~,\\
\label{eqn:obs3}
A^+_{CP} &=& \frac{{\rm Re}(\bar A- A)}{B_P} + \frac{{\rm Re}(\Delta \bar B
 - \Delta B)}{B_P}~~,\\
\label{eqn:obs4}
A^0_{CP} &=& -\frac{{\rm Re}(\bar A- A)}{B_P} + \frac{{\rm Re}(\Delta \bar B
 - \Delta B)}{B_P}~~.
\eea
The four asymmetries provide the following information:
\begin{itemize}
\item The $\Delta I = 0$ and $\Delta I = 1$ contributions in CP asymmetries
are separated by taking sums and differences,
\bea\label{ACP0}
A^{\Delta I=0}_{CP} & \equiv & \frac{1}{2}(A^+_{CP} + A^0_{CP}) = 
\frac{{\rm Re}(\Delta \bar B - \Delta B)}{B_P}~~,\\
\label{ACP1}
A^{\Delta I=1}_{CP} & \equiv & \frac{1}{2}(A^+_{CP} - A^0_{CP})  =
\frac{{\rm Re}(\bar A - A)}{B_P}~~.
\eea
\item ${\rm Re}A/B_P$ and ${\rm Re} \bar A/B_P$ may 
be separated by using information from $A^{\Delta I=1}_{CP}$ and $A_I$.
\item $\Delta S$ is governed by an {\it imaginary} part of a combination of 
$\Delta I = 0$ and $\Delta I = 1$ terms which cannot be separated in $B$ decays. 
Such a separation is possible in $B_s$ decays to pairs of isospin-reflected
decays, e.g. $B_s\to K^+K^-, K_SK_S$ or $B_s\to K^{*+}K^{*-}, K^{*0}\bar K^{*0}$,
where $2\beta$ in the definition of $\Delta S$ (\ref{DeltaS}) is now replaced by the 
small phase of $B_s$-$\bar B_s$ mixing.
\end{itemize}

One may take one step further under the assumption that strong phases associated 
with NP amplitudes are small relative to those of the SM and can be 
neglected.\cite{Datta:2004re} This assumption, which must be confronted 
by data, is reasonable because rescattering from a leading $b\to sc\bar c$ amplitude 
is likely the main source of strong phases, while rescattering from a smaller 
$b\to sq\bar q$ NP amplitude is then a second-order effect. 
In the convention (\ref{Bconvention}), where the strong phase of $B_P$ is set equal 
to zero,  $\Delta B$ and $A$ have the same CP-conserving strong phase $\delta$, and involve 
CP-violating phases $\phi_B$ and $\phi_A$, respectively, 
\beq\label{DeltaB,A}
\Delta B = |\Delta B|e^{i\delta}e^{i\phi_B}~~,~~~~~
A = |A|e^{i\delta}e^{i\phi_A}~~.
\eeq
Since the four asymmetries (\ref{eqn:obs1})-(\ref{eqn:obs4}) are first order in small ratios of
amplitudes, one may take $B_P$ in their expression to be given by the square root
of $\Gamma_+$ or $\Gamma_0$, thereby neglecting second order terms.
These four observables can then be shown to 
determine $|A|, \phi_A$ and $|\Delta B|\sin\phi_B$.\cite{Gronau:2007ut}  The 
combination $|\Delta B| \cos \phi_B$ adds coherently to $B_P$ and cannot be fixed 
independently.

The amplitudes $\Delta B$ and  $A$ consist of process-dependent SM and potential 
NP contributions. 
Assuming that the former are calculable, either using methods based on QCD-factorization 
or by fitting within flavor SU(3) these and other $B$ decay rates and asymmetries, the
four asymmetries determine the magnitude and CP violating phase of a $\Delta I=1$ NP 
amplitude and the imaginary part of a $\Delta I=0$ NP amplitude. In certain cases, e.g., 
$B\to \phi K$ or $B\to\eta'K_S$, stringent upper bounds on SM contributions to $\Delta B$ 
and  $A$ may suffice if some  of the four measured
asymmetries are larger than permitted by these bounds. In the pair
$B^+\to K^+\pi^0, B^0\to K^0\pi^0$, the four measured asymmetries [using the predicted value
(\ref{ACPK0pi0})] are $A_I=0.087 \pm 0.038, A^{\Delta I=0}_{CP}=-0.047\pm 0.025, A^{\Delta I=1}_{CP}=0.094\pm 0.025, \Delta S=-0.35\pm 0.21$. Some reduction of errors is required
for a useful implementation of this method.

\medskip\noindent
{\bf Conclusion}:  There exists ample experimental evidence in pairs of isospin-reflected 
$b\to s$ penguin-dominated decays that potential NP amplitudes must be small. Assuming
that these amplitudes involve negligible strong phases, and assuming that small SM 
non-penguin contributions are calculable or can be strictly bounded, 
one may determine the magnitude and CP violating phase of a NP $\Delta I=1$ amplitude, 
and the imaginary  part of a NP $\Delta I=0$ amplitude in each pair of isospin-reflected decays.
 
 \subsection{Null or nearly-null tests}
 
 We have not discussed null tests of the CKM framework.\cite{Gershon:2006mt} 
Evidence for physics beyond the Standard Model may show-up as (small) nonzero 
asymmetries in processes where they are predicted to be extremely small in the CKM framework. 
A well-known example is $B^+\to \pi^+\pi^0$, where the CP asymmetry is expected to be 
a small fraction of a percent including EWP amplitudes.\cite{Buras:1998rb,Gronau:1998fn}
We have only discussed {\em exclusive hadronic} $B$ decays, where QCD calculations involve 
hadronic uncertainties. A more robust calculation exists for the direct CP asymmetry in 
{\em inclusive radiative} decays $B\to X_s\gamma$, found to be smaller
than one percent.\cite{Soares:1991te} The current upper limit on this asymmetry is at
least an order of magnitude larger.\cite{Aubert:2004hq} 

Time-dependent asymmetries in radiative decays $B^0\to K_S\pi^0\gamma$, for 
a $K_S\pi^0$ invariant-mass in the $K^*$ region and for a larger invariant-mass range 
including this region, are interesting because they test 
the photon helicity, predicted to be dominantly right-handed in $B^0$ decays and 
left-handed in $\bar B^0$ decays.\cite{Atwood:1997zr,Atwood:2004jj} The asymmetry, 
suppressed by $m_s/m_b$, is expected to be several percent in the SM, and can be
very large in extensions where spin-flip is allowed in $b\to s\gamma$. While 
dimensional arguments seem to indicate a possible larger asymmetry in the SM, 
of order $\Lambda_{\rm QCD}/m_b\sim 10\%$,\cite{Grinstein:2004uu} calculations 
using perturbative QCD\,\cite{Matsumori:2005ax} and QCD factorization\,\cite{Ball:2006cv} 
find asymmetries of a few percent. The current averaged values, for the $K^*$ region and
for a larger invariant-mass range including this region, are 
$S((K_S\pi^0)_{K^*}\gamma)=-0.28\pm 0.26$ and $S(K_S\pi^0\gamma)=
-0.09\pm 0.24$.\cite{HFAG,Aubert:2005bu}  These measurements must be improved 
in order to become sensitive to the level predicted in the SM, or to provide evidence for physics 
beyond the SM.

\section{Summary}

The Standard Model passed with great success numerous tests in the flavor sector, including
a variety of measurements of CP asymmetries related to the CKM phases $\beta$ and $\gamma$. 
Small potential New Physics corrections may occur in $\Delta S=0$ and $|\Delta S|=1$ penguin amplitudes, affecting the extraction of $\gamma$ and modifying CP-violating and
isospin-dependent asymmetries in $|\Delta S|=1$ $B^0$ decays and 
isospin-related $B^+$ decays.  Higher precision than achieved 
so far is required for claiming evidence for such effects and for sorting out their isospin structure.

Similar studies can be performed with $B_s$ mesons produced at hadron colliders and at 
$e^+e^-$ colliders running at the $\Upsilon(5S)$ resonance. Time-dependence
in $B_s\to D^-_sK^+$ and $B_s\to J/\psi\phi$ or $B_s\to J/\psi\eta$ measures $\gamma$ 
and  the small phase of the $B_s$-$\bar B_s$ mixing amplitude.\cite{Aleksan:1991nh} 
Comparing time-dependence and angular analysis in $B_s\to J/\psi\phi$  with $b\to s$ 
penguin-dominated processes including $B_s\to \phi\phi, B_s\to K^{*+}K^{*-}, 
B_s\to K^{*0}\bar K^{*0}$ provides a methodic search for potential NP effects. 
Work on $B_s$ decays has just begun at the Tevatron.\cite{Paulini:2007mf} One is 
looking forward to first results from the LHC.

\section*{Acknowledgments}
I am grateful to numerous collaborators, in particular to Jonathan Rosner whose
collaboration continued without interruption for many years.
This work was supported in part by the Israel Science Foundation
under Grant No.\ 1052/04 and by the German-Israeli Foundation under
Grant No.\ I-781-55.14/2003.


\begin{thebibliography}{0}
\bibitem{References} This review, which is only 27 page long (the number of 
Hebrew alphabet letters) includes 120 references, as a Jewish blessing says 
``May you live to be 120!" It is too short to include other hundreds or thousands 
of relevant papers. I apologize to their many authors.
\end{thebibliography}
\end{document}